\documentclass[]{elsarticle}
\usepackage{amsmath}   
\usepackage{graphicx}   
\begin{document}
 
\title{Fine-sorting One-dimensional Particle-In-Cell Algorithm with Monte-Carlo Collisions on a Graphics Processing Unit}
 
\author[rub]{Philipp Mertmann\corref{cor1}}
\ead{mertmann@aept.rub.de}
\address[rub]{Institute for Electrical Engineering and Plasma
  Technology, Ruhr-Universit\"at Bochum, Universit\"atsstr. 150, 44801~Bochum, Germany, Tel:~++49~(0)234~3228062, Fax:~++49~(0)234~3214230}
\author[pu]{Denis Eremin}
\author[pu]{Thomas Mussenbrock}
\author[pu]{Ralf Peter Brinkmann}
\address[pu]{Institute for Theoretical Electrical Engineering,
  Ruhr-Universit\"at Bochum}
\author[rub]{Peter Awakowicz}

\cortext[cor1]{Corresponding author}

\begin{keyword}
GPU \sep PIC \sep Particle-In-Cell \sep sorting algorithm \sep CUDA
\end{keyword}

\begin{abstract}
Particle-in-cell (PIC) simulations with Monte-Carlo collisions are used in plasma science to explore a variety of kinetic effects. One major problem is the long run-time of such simulations. Even on modern computer systems, PIC codes take a considerable amount of time for convergence. Most of the computations can be massively parallelized, since particles behave independently of each other within one time step. Current graphics processing units (GPUs) offer an attractive means for execution of the parallelized code. In this contribution we show a one-dimensional PIC code running on Nvidia\textsuperscript{\textregistered} GPUs using the CUDA\texttrademark\ environment. A distinctive feature of the code is that size of the cells that the code uses to sort the particles with respect to their coordinates is comparable to size of the grid cells used for discretization of the electric field. Hence, we call the corresponding algorithm ``fine-sorting''. Implementation details and optimization of the code are discussed and the speed-up compared to classical CPU approaches is computed.
\end{abstract}

\maketitle

\section{Introduction}
In the past years, the development of central processing units (CPUs) for consumers changed from just increasing the clock frequency and integrating special functions into the core to the design of processors with multiple cores. Accordingly, the use of parallel computing becomes more and more important, even though most computer programs presently do not take advantage of this possibility. One branch that optimized the parallelization perfectly and also optimized their hardware for that purpose is the manufacturing industry of graphics cards. Modern 3-D applications allow for the parallel computation of many independent values, such as pixels on a display. 

Number of scientific efforts that take advantage of the computational power of graphics processing units (GPU) is growing. Recently, many publications were made where calculations benefit from the speed of graphics processing units \cite{hoomd,badal:4878,kersevan:1017,Anderson2007298,DBD_paper_1,doi:10.1021/ct900275y,doi:10.1021/ct9005079}. Main convenience of this approach is a high speed multi core processing unit that is comparatively low priced.

Low pressure plasmas play a major role in a wide range of industrial applications, including, among other things, semi-conductor processing, surface coating or sterilization processes. Characteristics of these gas discharges, such as pressure, input power, reactor sizes, length scales of particle collisions or the typical length of electric field screening (Debye-length) span several orders of magnitude, depending on the actual application. This makes modeling extremely difficult, because many assumptions or simplifications of a specific model are usually only applicable to a particular case. 

Fluid models, which consider velocity distributions functions (mostly Maxwellian) become improper under certain conditions, due to their nature of using integral quantities and principles like diffusion. PIC simulations \cite{birdsall_book, hockney_book, tajima_book} trace particles (as electrons and ions) on their way inside a plasma, being impacted by collisions, the electric field and the walls of the reactor. Thereby, distribution functions of particles are approximated by PIC simulations. By averaging (time or space integrations) over certain quantities, macroscopic and measurable values such as particle density, flux and current can be calculated. Since only a few assumptions are made, particle-in-cell (PIC) simulations retain most of the fundamental, nonlinear effects in a plasma.

In this article we present a PIC code parallelized for execution on a graphics card, using Nvidia's CUDA environment in C. In the second section we give a short introduction into particle-in-cell simulations. In section \ref{sec:implementation_details} the algorithm is explained in detail. In section \ref{sec:results} we run the code under different conditions and obtain parameters for an optimized speed-up.

\section{Particle-In-Cell Basics}
Particle-in-cell simulations follow the movement of charged particles inside a plasma. Actually, every simulated (super) particle represents a certain amount of real particles. Particle locations and velocities are defined in continuum space, whereas charge density, electric potential and field are defined on a spatial grid of size $N_{\textrm{grid}}$. This work examines one spatial and three velocity components of the particles. 

\begin{figure}
  \begin{center}
    \includegraphics[width=\textwidth]{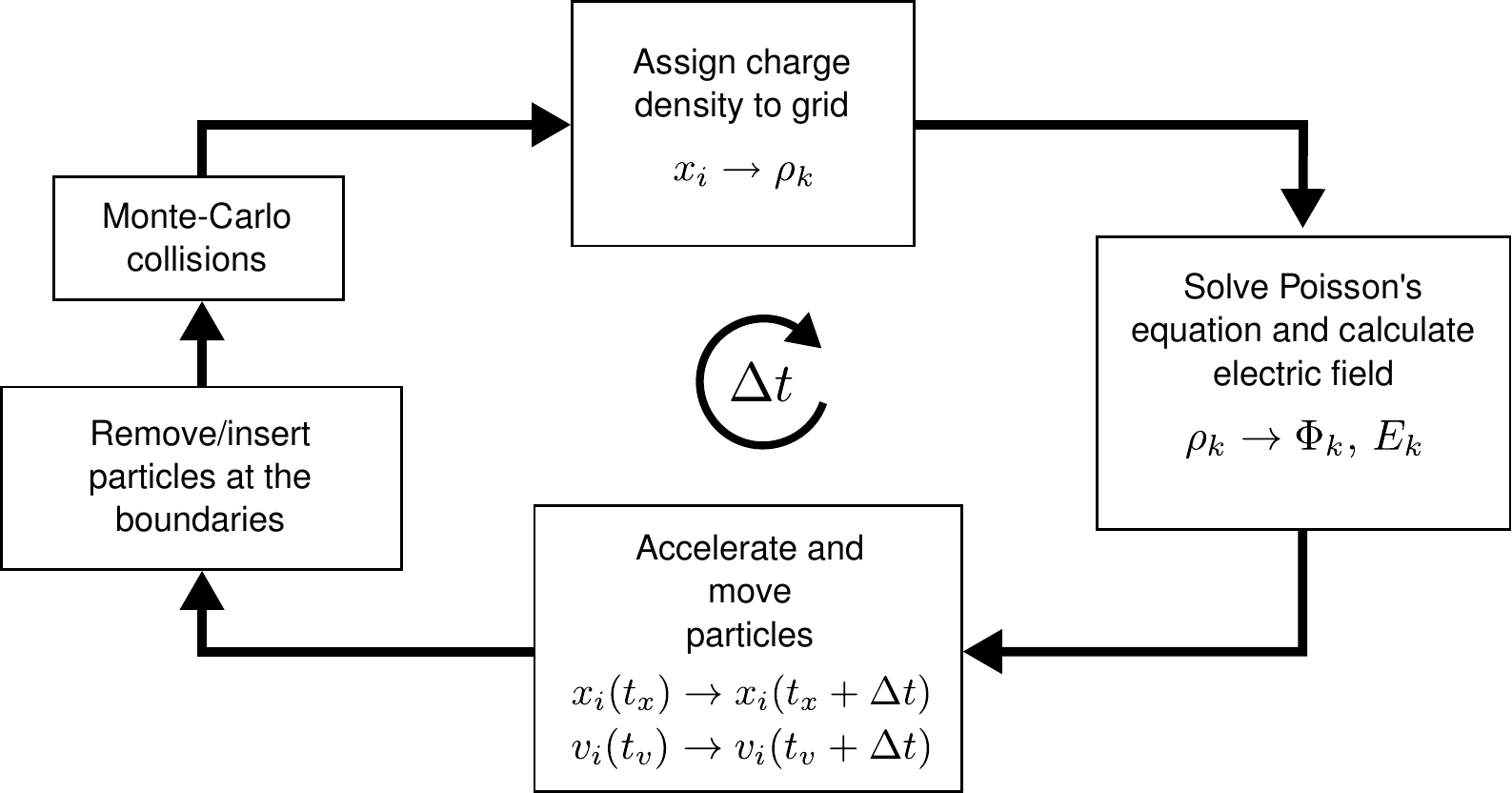}
  \end{center}
  \caption{Computing sequence of the particle-in-cell cycle with Monte-Carlo collisions.}
  \label{fig:pic_cycle}
\end{figure}
Figure \ref{fig:pic_cycle} shows the modules of the PIC algorithm, that are passed through every finite time step $\Delta t$. Beginning at the top, charge density is calculated by weighting particles on the grid, using a linear interpolation (first order) scheme in this work. After that, Poisson's equation is solved and the electric field is calculated. Field values on the grid point are interpolated to particle positions and then particles accelerate and move inside the simulation box. Particles that left the finite box are removed from the simulation every $\Delta t$. If secondary electrons or electron reflection are featured, electrons are inserted at the boundaries. Now collisions take place, handled by a Monte-Carlo module and the loop is closed.

To render an actual plasma behavior, some conditions have to be satisfied. The finite grid spacing (distance between two grid points) has to be smaller than the Debye-length to resolve field effects on the correct length scale. Additionally, time step size has to be small enough to resolve the highest frequency processes in the system, which are oscillations with electron plasma frequency. Recently it was shown that the number of particles inside a Debye-sphere has to be larger than expected until then to reduce or avoid spurious numerical effects \cite{turner:033506}. All this leads to an enormous amount of computational power, needed by such simulations.

\section{Implementation Details} \label{sec:implementation_details}
\subsection{GPU Programming with CUDA}
Nvidia's \textit{Compute Unified Device Architecture} (CUDA) is an architecture for parallel computations on graphics processing units. Programs are written in the programming language C as regular C-code with some extensions to provide access to the GPU \cite{cuda_programming_guide}. 

Graphics cards have a number of parallel multiprocessors, depending on the generation and model of the GPU. A multiprocessor features a certain amount of processor cores, each having its own fast register memory, but the cores of a multiprocessor also share the on-chip \textit{shared memory} space. All different multiprocessors have access to the large \textit{global memory}. Additionally, there are two cached read-only memory spaces, \textit{constant memory} and \textit{texture memory}.

Special functions, called \textit{kernels}, run on the GPU in parallel, using thousands or millions of independent \textit{threads}. Threads are grouped into \textit{thread blocks}, whose size is chosen by the programmer. All threads of such a block run on the same physical multiprocessor, thus having access to the same shared memory space. Constant and texture memory are read-only spaces for threads and can be exclusively filled with data from outside a kernel by special functions.

\subsection{Data Structure}
\begin{figure}
  \begin{center}
    \includegraphics[width=\textwidth]{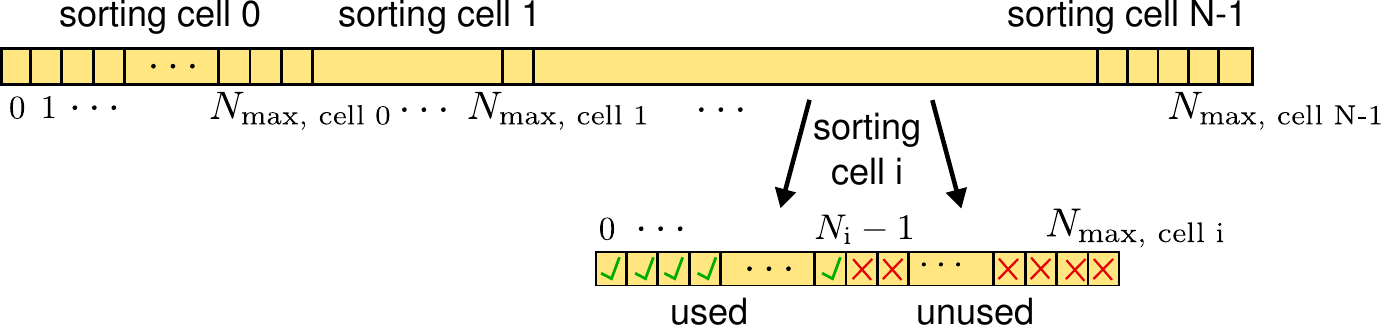}
  \end{center}
  \caption{Fine-sorted particle array. Each sorting cell $i$ has a fixed size $N_\textrm{max}$ and is filled with particle data up to a certain number $N_i$. Information about the current status $N_i$ of a sorting cell has to be stored in a separate (integer) array.}
  \label{fig:particle_array}
\end{figure}

Particle data is stored in a float4 array, using the x, y and z component for velocities and the w component for the location of a particle. Each particle species (such as electrons or ions) uses its own array. All arrays are divided into sorting cells, each of the same fixed size (figure \ref{fig:particle_array}). Each sorting cell in GPU memory belongs to a certain region in configuration space, accordingly all particles within a sorting cell are close to each other. Information on how many particles $N_i$ currently reside in a sorting cell is stored in an integer array, one integer for each cell. All sorting cells therefore are partially filled with the currently residing particles and the rest is free for new particles entering the cell.

Such sorting cells in memory can contain multiple grid points of the particle-in-cell spatial grid, that defines the values of charge density, potential and field. The number of grid points per sorting cell $N_{\textrm{gc}}$ can be configured by the user at the beginning of a simulation and influences the speed.

\begin{figure}
  \begin{center}
    \includegraphics[width=\textwidth]{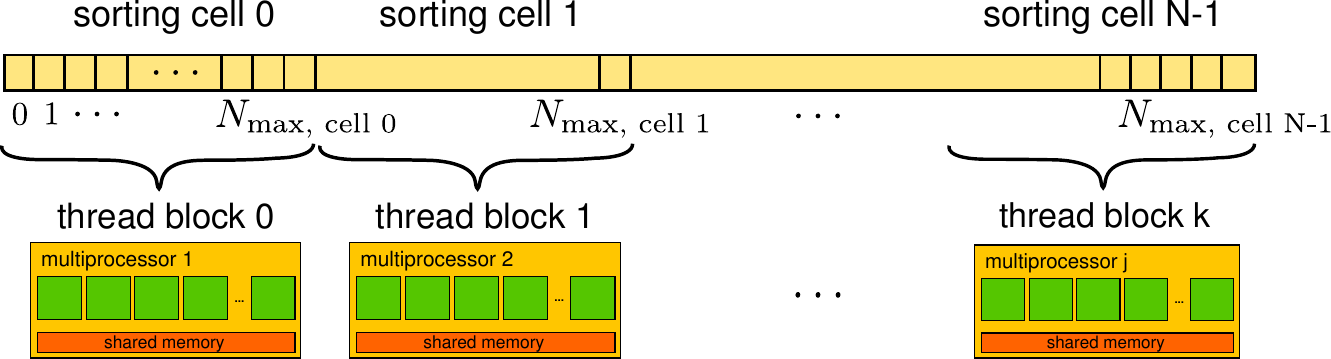}
  \end{center}
  \caption{Calling a kernel function in the fine-sorted algorithm. Each cell is handled by a single thread block and thus by a single multiprocessor.}
  \label{fig:kernel_calls}
\end{figure}

If a kernel is called, each sorting cell is handled by a single block of threads (and thus a single multiprocessor, figure \ref{fig:kernel_calls}). Using the information from the integer array (current number of particles inside a sorting cell), threads of a block can treat (e.g. push) all valid particles of a sorting cell and leave the remaining parts of the float4 array unaffected. 

The advantage of a fine-sorting code is, that this kind of code allows for a straightforward and fast implementation of nonlinear collision processes (i.e. Coulomb collisions) through the binary collisions, which is difficult otherwise. In this case, size of the sorting cells must not exceed the Debye-length. Additionally, other modules of the PIC cycle benefit from the fine-sorting approach, as stated in the following. A good overview of advantages and drawbacks of different ways of GPU implementations can be found in \cite{Stantchev20081339}. It is worth noting that unlike the codes on CPUs, such as \cite{Tskhakaya2007829}, we did not observe any significant improvement of the code run-time compared to an alternative GPU algorithm, where particle data is unsorted, as will be described in our future work.

\subsection{Normalization}
Equations are normalized to minimize the amount of additional factors and thereby reduce the computational effort by the code. Finite difference integration of the equations of motion lead to multiplications with the time step $\Delta t$, so normalizing time $t$ on this time step leads to the simplest case of a multiplication with $1$
\begin{equation*}
  t \rightarrow \Delta t \cdot \tilde{t}
\end{equation*}

Determination of the nearest grid point $p_i$ of a particle is an important part of all grid dependent calculations. Normalizing the spatial coordinate $x$ on the lattice spacing $\Delta x$ 
\begin{equation*}
  x \rightarrow \Delta x \cdot \tilde{x}
\end{equation*}
involves a single type casting operation (float to integer) to calculate $p_i$.

To simplify the charge density assignment, charge density $\rho$ is normalized on the charge density of a single super-particle
\begin{equation*}
 \rho \rightarrow \dfrac{Q}{A_{\textrm{e}}\,\Delta x} \cdot \tilde{\rho},
\end{equation*}
with $Q$, the charge of super-particle and $A_{\textrm{e}}$ the electrode area of the discharge.

Electric potential $\Phi$ is normalized to ease Poisson's equation
\begin{equation*}
 \Phi \rightarrow \dfrac{Q\Delta x}{\epsilon_0\,A_{\textrm{e}}} \cdot  \tilde{\Phi},
\end{equation*}
with $\epsilon_0$, the the vacuum permittivity. Electric field is normalized on
\begin{equation*}
E \rightarrow \dfrac{Q}{\epsilon_0\,A_{\textrm{e}}} \cdot \tilde{E} 
\end{equation*}
and velocities are normalized on
\begin{equation*}
 v \rightarrow \dfrac{\Delta x}{\Delta t} \cdot \tilde{v}.
\end{equation*}
Skipping all tildes for normalized values, the equations read
\begin{align}
 & \dfrac{\partial^2 \Phi}{\partial x^2} = -\rho(r,z) \label{eq:poisson_normalized} \\
 & E = -\dfrac{\partial \Phi}{\partial x} \\
 & \dfrac{\partial v}{\partial t} =  \dfrac{Q^2 \Delta t^2}{\epsilon_0\,\Delta x M \,A_{\textrm{e}}} E = \kappa E \label{eq:push_velocity} \\
 &  \dfrac{\partial x}{\partial t} = v \label{eq:push_location} 
\end{align}
with $M$ the mass of a super-particle.

\subsection{Charge Density Assignment}
\begin{figure}
  \begin{center}
    \includegraphics[width=\textwidth]{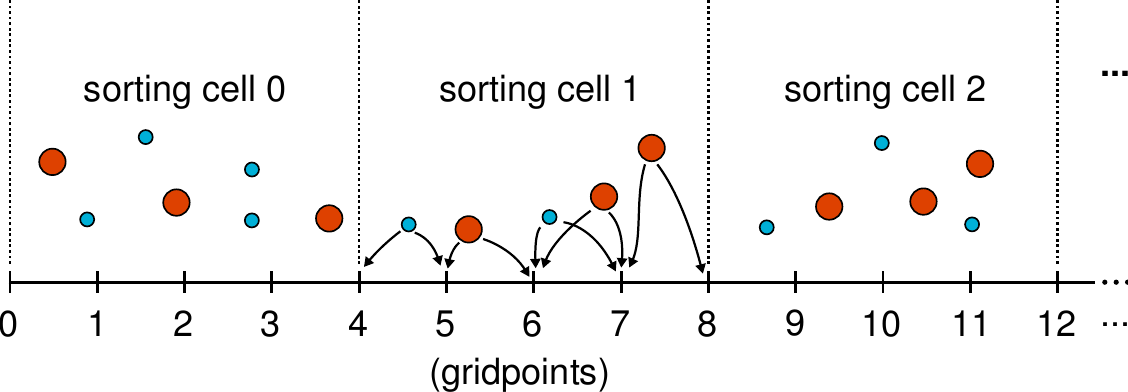}
  \end{center}
  \caption{Ions (red) and electrons (blue) of one sorting cell only contribute to the charge density of grid points that belong to their cell (linear weighting). Each sorting cell contains 5 grid points in this case.}
  \label{fig:super_cell}
\end{figure}

Before the kernel is started, the number of particles in each sorting cell $N_i$ is copied to constant memory. All threads of a block treat particles of a single sorting cell. Therefore particles of one block only contribute to a small area in the charge density array, namely $N_{\textrm{gc}}$ grid points belonging to that sorting cell (figure \ref{fig:super_cell}).

Threads allocate a local array of $N_{\textrm{gc}}$ floats in register memory and initialize the array with zeros. Afterwards each thread starts loading a particle to register memory, adding its charge to the local charge density array. The $i$-th thread of each block thereby executes the $i$-th particle of the sorting cell. If there are more particles than the size of the block, this procedure is repeated in strides equal to the block size as many times as needed.

\begin{figure}
  \begin{center}
    \includegraphics[width=\textwidth]{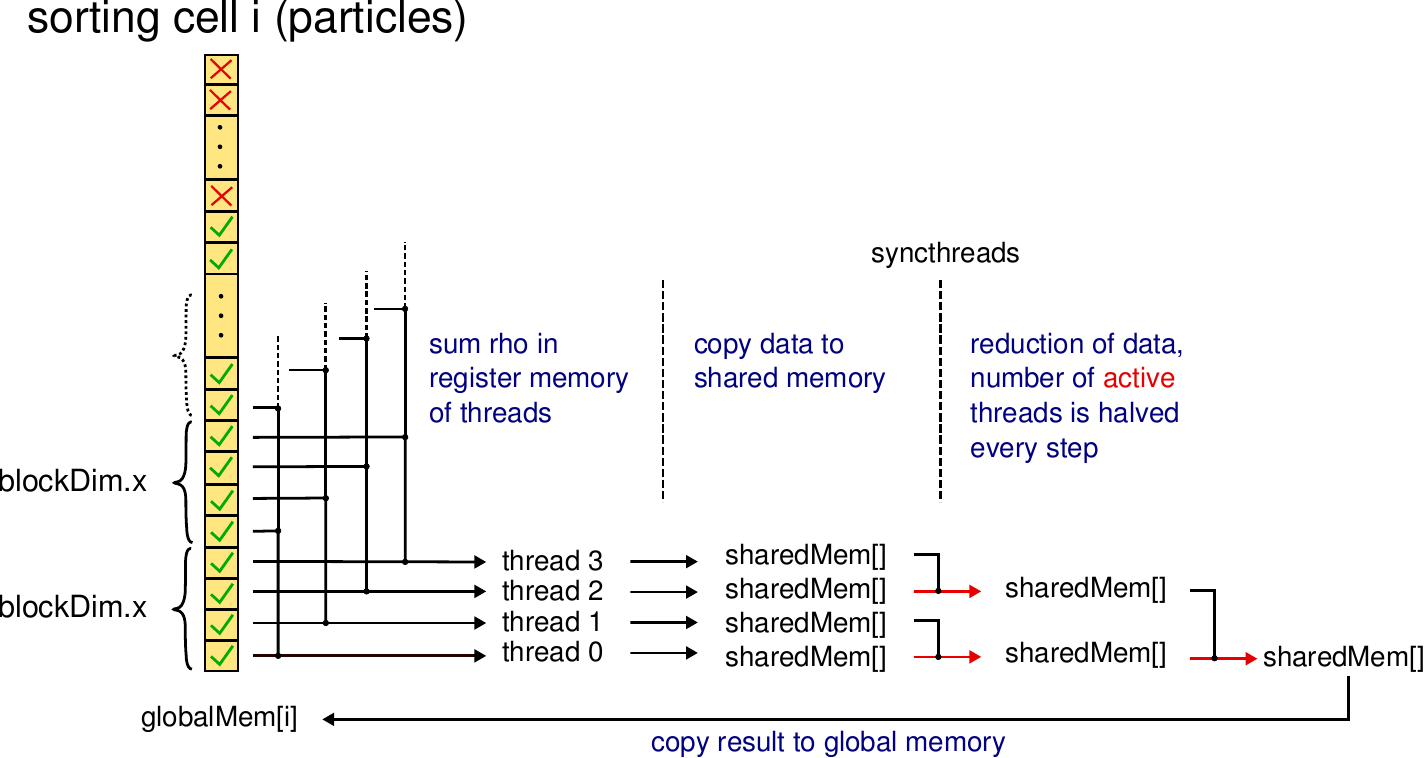}
  \end{center}
  \caption{Charge density calculation. Threads sum rho in register memory, exemplary shown for only four threads and a single sorting cell $i$. The reduction is done using shared memory, the last active thread stores the result in global memory.}
  \label{fig:rho_calc}
\end{figure}

Once all particles are done, a reduction in shared memory starts. Each thread copies its whole local array to a shared memory array. Now half of the threads becomes inactive. Each active thread adds the charge density of an inactive thread to its own shared array. Again and again half of the threads becomes inactive, until only a single thread is left. This thread copies the reduced charge density of the current cell to global memory. The very left and right grid points of a sorting cell are used by the cell and the neighboring cell, respectively. To avoid global memory atomics, the right grid point of each sorting cell is stored separately in a float array in global memory and is added to the charge density in a subsequent kernel function. The whole procedure is shown in figure \ref{fig:rho_calc}. 

\subsection{Field Solver}
Discretizing Eq.(\ref{eq:poisson_normalized}) with finite differences leads to a tridiagonal matrix. A detailed description of including different boundary conditions can be found in \cite{160337}. Solving can be done very efficiently serially on the CPU, using the Thomas algorithm \cite{thomas_algorithm}, an optimized Gaussian elimination scheme for tridiagonal matrices. 

Since charge density is calculated on the GPU, the array must be copied to CPU memory, Poisson's equation is solved and the electric field is calculated. After this, the electric field array is copied to GPU constant memory. Despite the amount of memory transfer between CPU and GPU, this module takes only a small part of the run-time of the PIC cycle. For one million electrons and ions, respectively, and $800$ grid points for the field, data copy and solving takes about $3.55\%$ of the run-time of a cycle. Thus, a more complicated parallel approach that runs completely on the GPU does not seem to be worth it for a one-dimensional simulation. For large number of grid points or a two-dimensional approach, this is of course not the case.

\subsection{Particle Pusher}
In this kernel, the current number of particles in each sorting cell $N_i$ and the electric field are read from constant memory. Each thread loads a particle to register memory and updates the particle velocity by multiplying the interpolated field with $\kappa$ (Eq.(\ref{eq:push_velocity})). After this, the new particle location is calculated (Eq.(\ref{eq:push_location})), using the new velocity. This relates to the commonly used leap-frog algorithm, where location and velocity are separated in time by half of a time step \cite{hockney_book,birdsall_book}. 
\begin{align*}
v\left(t+0.5\right) - v\left(t-0.5\right) &= \kappa E\left(x\left(t\right)\right) \\
x\left(t+1\right) - x\left(t\right) &= v\left(t+0.5\right) 
\end{align*}
The result is stored in global memory. Just as for charge density assignment, threads will load additional particles, if the number of particles exceeds the block size.

\subsection{Fine-Sorting Algorithm}
After each pushing, some particles reside in incorrect sorting cells. Since the Courant-condition \cite{hockney_book,birdsall_book} needs to be satisfied, this is only a small fraction of particles.
\begin{equation}
 v_{\textrm{e}} < \dfrac{\Delta x}{\Delta t} < N_{\textrm{gc}} \dfrac{\Delta x}{\Delta t} \label{eq:courant}
\end{equation}
$v_{\textrm{e}}$ is the thermal velocity of electrons, so inequality (\ref{eq:courant}) basically means, that only a small number of particles leaves its sorting cell during one time step. Accordingly, sorting is optimized for a ``nearly sorted'' particle array. 

\begin{figure}
  \begin{center}
    \includegraphics[width=\textwidth]{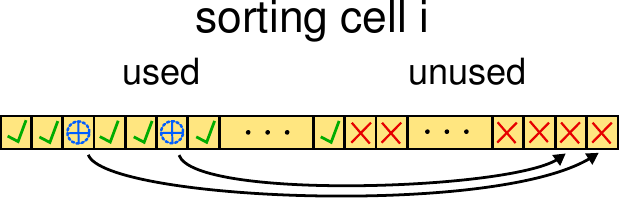}
  \end{center}
  \caption{Step 1a of the sorting algorithm for sorting cell $i$. Rejected particles (blue symbols) move to free positions (red crosses) at the end of sorting cell $i$.}
  \label{fig:sorting_step1a}
\end{figure}

In a first kernel, threads copy incorrect particles to free positions at the end of their current sorting cell. This memory space serves as a buffer for sorting particles (see Fig. (\ref{fig:sorting_step1a})).
Since particles of one sorting cell are checked in parallel by threads of a single block, all threads have access to the same shared memory. Hence, a counter $N_{\textrm{index, }i}$ in shared memory can provide indices for the new particle positions at the end of the sorting cell. If a thread spots an invalid particle, it decreases $N_{\textrm{index, }i}$ by one, using the \textit{atomicSub} function and thereby gets the new index. 

\begin{figure}
  \begin{center}
    \includegraphics[width=\textwidth]{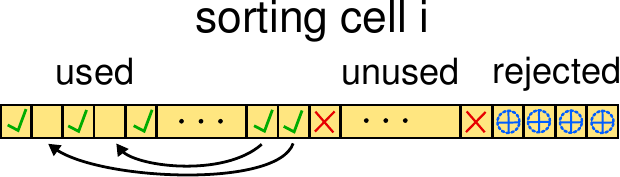}
  \end{center}
  \caption{Step 1b of the sorting algorithm for sorting cell $i$. Gaps are closed with valid particles from current sorting cell $i$.}
  \label{fig:sorting_step1b}
\end{figure}

After this, each sorting cell has gaps (blue symbols in Fig.(\ref{fig:sorting_step1a})) which have to be filled with valid particles. Threads scan particles a second time and fill the gaps with information from the last valid particle in each sorting cell, respectively (Fig.(\ref{fig:sorting_step1b})). 
Therefore a second counter in shared memory allocates correct particle's indices at position $N_{i}$, $N_{i}-1$ and so on. These two steps can be included in the particle pusher, to re-use particle data efficiently and thereby reduce the memory transfer from global memory. The number of rejected particles in each sorting cell and the new number of valid particles $N_{i}$ are stored in integer arrays in global memory.

The second step of sorting starts in a new kernel. 
\begin{figure}
  \begin{center}
    \includegraphics[width=\textwidth]{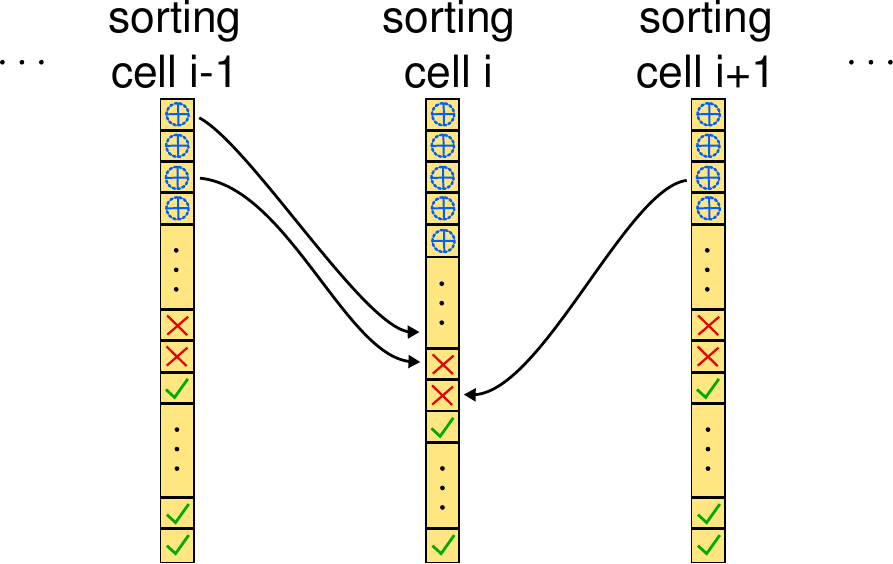}
  \end{center}
  \caption{Step 2 of the sorting algorithm. Threads of block $i$ copy rejected particles from the neighboring sorting cells to cell $i$.}
  \label{fig:sorting_step2}
\end{figure}
Again, a thread block is started for each sorting cell $i$. Now, half of the threads checks the rejected particles of the left neighbor $i-1$, the other half checks the right neighbor $i+1$  (figure \ref{fig:sorting_step2}). A counter in shared memory defines the new particle index in sorting cell $i$ and is incremented with the \textit{atomicAdd} function. Particles from the neighboring sorting cells are just read, so no atomic operations are needed in global memory. The updated number of valid particles is stored in global memory.

\begin{figure}
  \begin{center}
    \includegraphics[width=\textwidth]{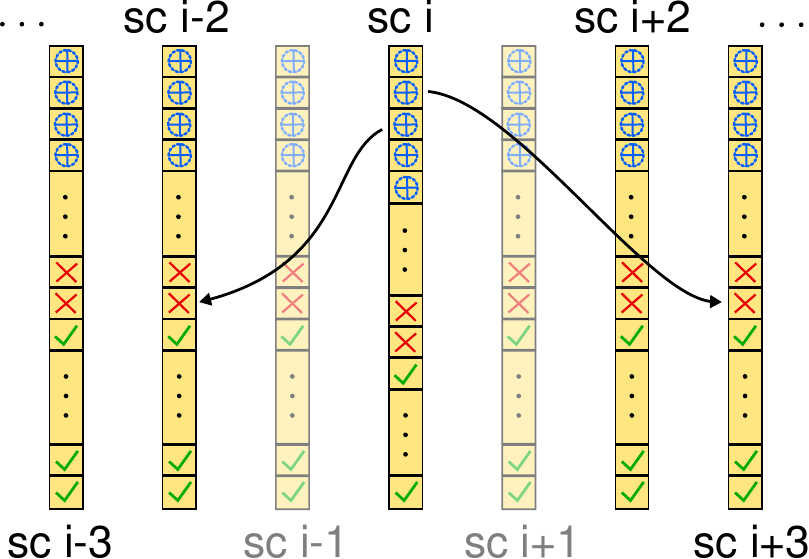}
  \end{center}
  \caption{Step 3 of the sorting algorithm. Threads of block $i$ copy rejected particles from their own sorting cell (sc) to other cells using atomic functions in global memory for the index counter.}
  \label{fig:sorting_step3}
\end{figure}

The last kernel of the sorting algorithm has to identify rejected particles, that are not sorted into the correct sorting cells yet. These are rare particles that moved farther than one sorting cell. Threads of a thread block now check rejected particles of its own sorting cell. If such a fast particle is located, its information is copied to the correct sorting cell (Fig.(\ref{fig:sorting_step3})).
For assigning indices to the particle, a counter in global memory is needed. This is the sole exception for the use of atomic functions in global memory. Since threads from all blocks access the same indices this step has to be done in global instead of shared memory. Again, the number of particles is stored in global memory. After this, all particles are arranged in correct sorting cells.

\subsection{Monte-Carlo Collisions}
Monte-Carlo collisions are treated by a modified null collision method. A detailed description of the standard algorithm can be found in \cite{verboncoeur_review, Birdsall_paper}. 

Here, null collision frequency $\nu_{\textrm{0}}$ is calculated once in the usual way, but instead of picking 
\begin{equation}
  N_{0} = \left( 1 - \textrm{e}^{-\Delta t \, \nu_\textrm{0}} \right) N_{\textrm{p}} =  P_{\textrm{0}} N_{\textrm{p}} \label{eq:null_collision}
\end{equation}
particles randomly from the total $N_{\textrm{p}}$ number of particles, a random number $p$ is drawn for every particle. This number is compared with $P_{\textrm{0}}$, the null collision probability. For $p < P_{\textrm{0}}$ the regular null collision algorithm is started, that is calculating collision cross sections and probabilities for the current particle. At an average $N_{0}$ particles run through the standard null collision algorithm, which satisfies equation \ref{eq:null_collision}. Using this modified null collision method, global memory atomics can be avoided, because every particle is just treated by one individual thread. If a particle is created due to ionization, it is injected into the sorting cell of the colliding electron. Thus, a counter in shared memory can allocate indices for those new particles.

Consequently, each thread needs its own random number generator (RNG). In principal, every RNG that uses only a single seed is sufficient, because memory access for initializing the RNG has to be small for a fast implementation. In this work we use a 3-xor-shift (11,7,12) generator \cite{Marsaglia:2003:JSSOBK:v08i14}, which can produce random numbers very quickly on the GPU. An unsigned integer array allocates seeds. Each thread loads its own seed from global memory, generates a certain number of random values and stores the last seed back to global memory.

CPU simulations usually use look-up tables for collision cross sections. On the GPU, simple functions, using only products and sums, can be much faster than scattered reads from memory. Therefore, cross sections for argon are calculated directly, using approximations of measured values \cite{Phelps_ionen_cross_sections, phelps:747, argon_electronen_cross_sections}.

Even though this implementation creates lots of branches, it is much faster than CPU collision handling.

\section{Optimizing the Parameters and Speed-up Measurements} \label{sec:results}
\subsection{Variation of Grid Points per Sorting Cell}
The algorithm sorts particles into sorting cells, which can contain any number of grid points of the charge density or rather the electric field array (figure \ref{fig:super_cell}). At the smallest possible size, there are just two grid points per sorting cell, one on the left, one on the right. The number of particles inside each sorting cell is low and all kernels have to be called for a large number of cells. It is obvious, that overhead due to calling functions and kernels slows down this approach. On the other hand, the reduction algorithm is not suitable for a large number of grid points per sorting cell $N_{\textrm{gc}}$ and also the amount of register memory can be limiting. Thus, the GPU may run slower than for smaller sorting cells. Also, there is an upper boundary on $N_{\textrm{gc}}$ governed by the requirement that sorting cell size does not exceed the Debye-length, in case one wants to implement Coulomb collisions. 

Between these to extremes one can suspect an optimized version. 
\begin{figure}
  \begin{center}
    \includegraphics[width=\textwidth]{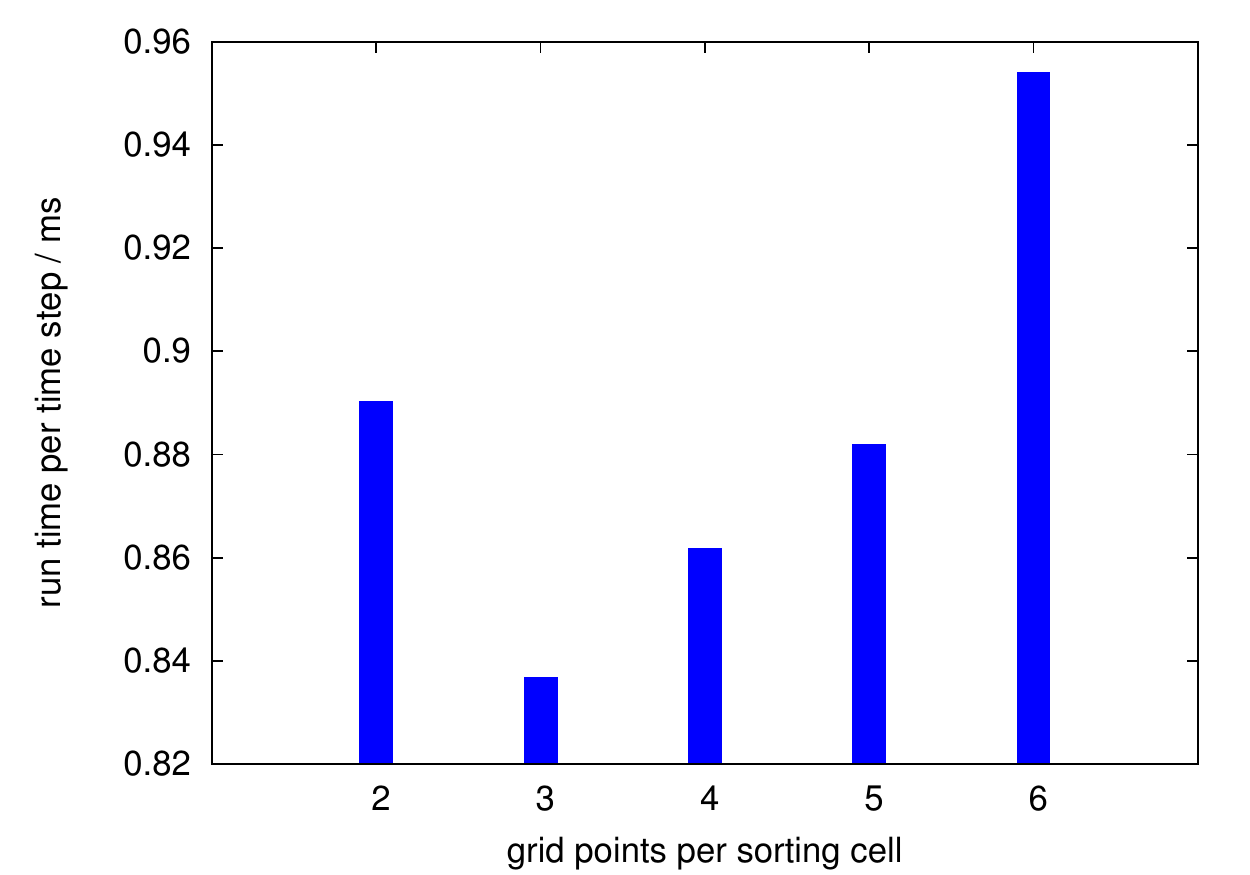}
  \end{center}
  \caption{Execution time of a single cycle for different numbers of grid points per sorting cell $N_{\textrm{gc}}$. Simulation runs with 500,000 ions and electrons, respectively at a pressure of 10~Pa.}
  \label{fig:gppsc}
\end{figure}
Figure \ref{fig:gppsc} shows the run-time of a single cycle of the simulation as a function of $N_{\textrm{gc}}$. In fact, a minimum can be found for $N_{\textrm{gc}}=3$. Note, that for some cases the minimum is found at $N_{\textrm{gc}}=4$, depending on the generation of GPU and all input values of the simulation.

\subsection{Variation of the Block Size}
\begin{figure}
  \begin{center}
    \includegraphics[width=\textwidth]{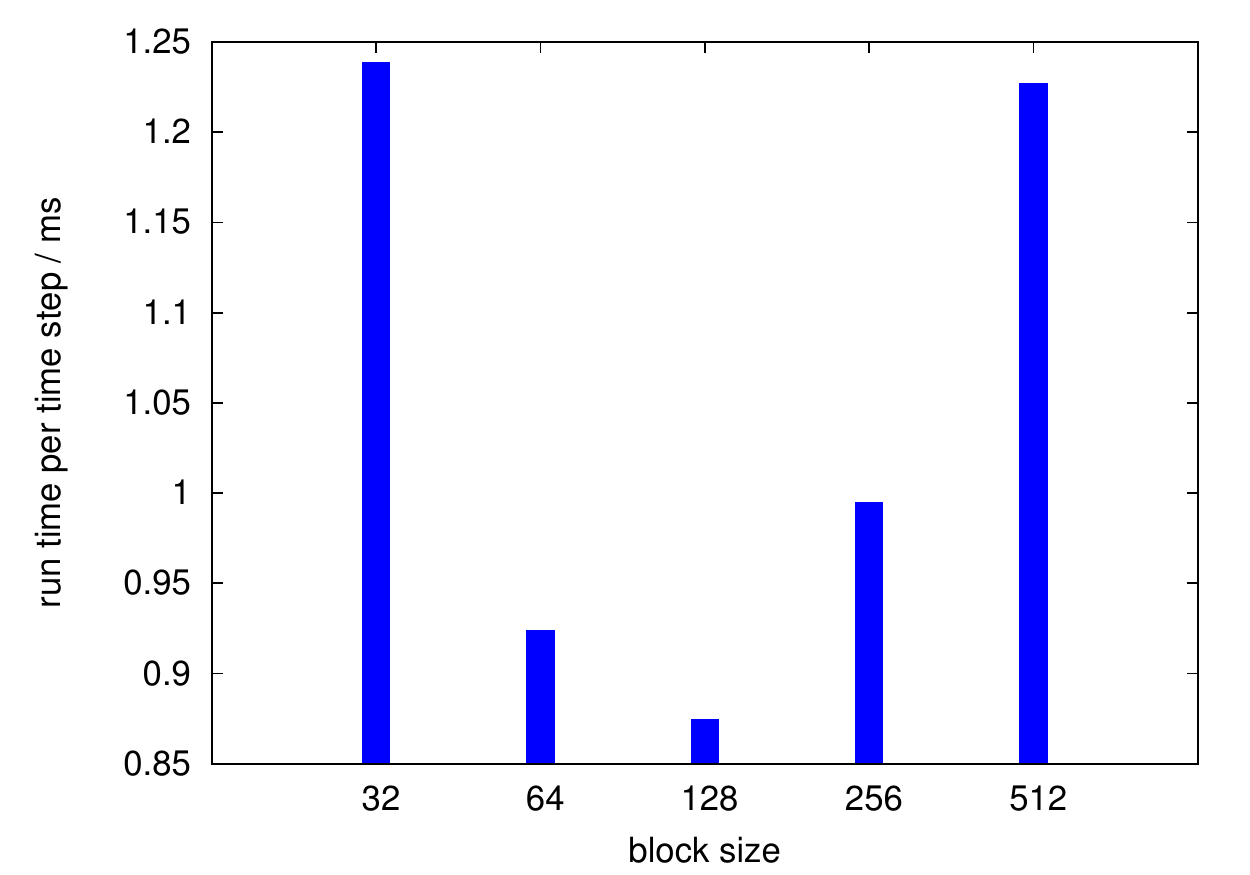}
  \end{center}
  \caption{Execution time of a single cycle for different block sizes. Simulation runs with 500,000 ions and electrons, respectively at a pressure of 10~Pa.}
  \label{fig:block_size}
\end{figure}

CUDA allows for choosing the size of blocks with a maximum of 512 or 1024 threads per block, for compute capability 1.2 and 2.0 hardware, respectively. In many applications, large blocks with many parallel threads are the best choice. Since CUDA hides memory access latencies of a thread by handling other threads meanwhile, small blocks are usually not the best option. For different inputs a block size of 128 had the best performance. Figure \ref{fig:block_size} shows the run-time over the block size for a system of 500,000 ions and electrons, respectively. 
These results were obtained with the same block size for all different kernels, but also every single module showed a similar behavior.

\subsection{Electrical Grid Size Dependence}
\begin{figure}
  \begin{center}
    \includegraphics[width=\textwidth]{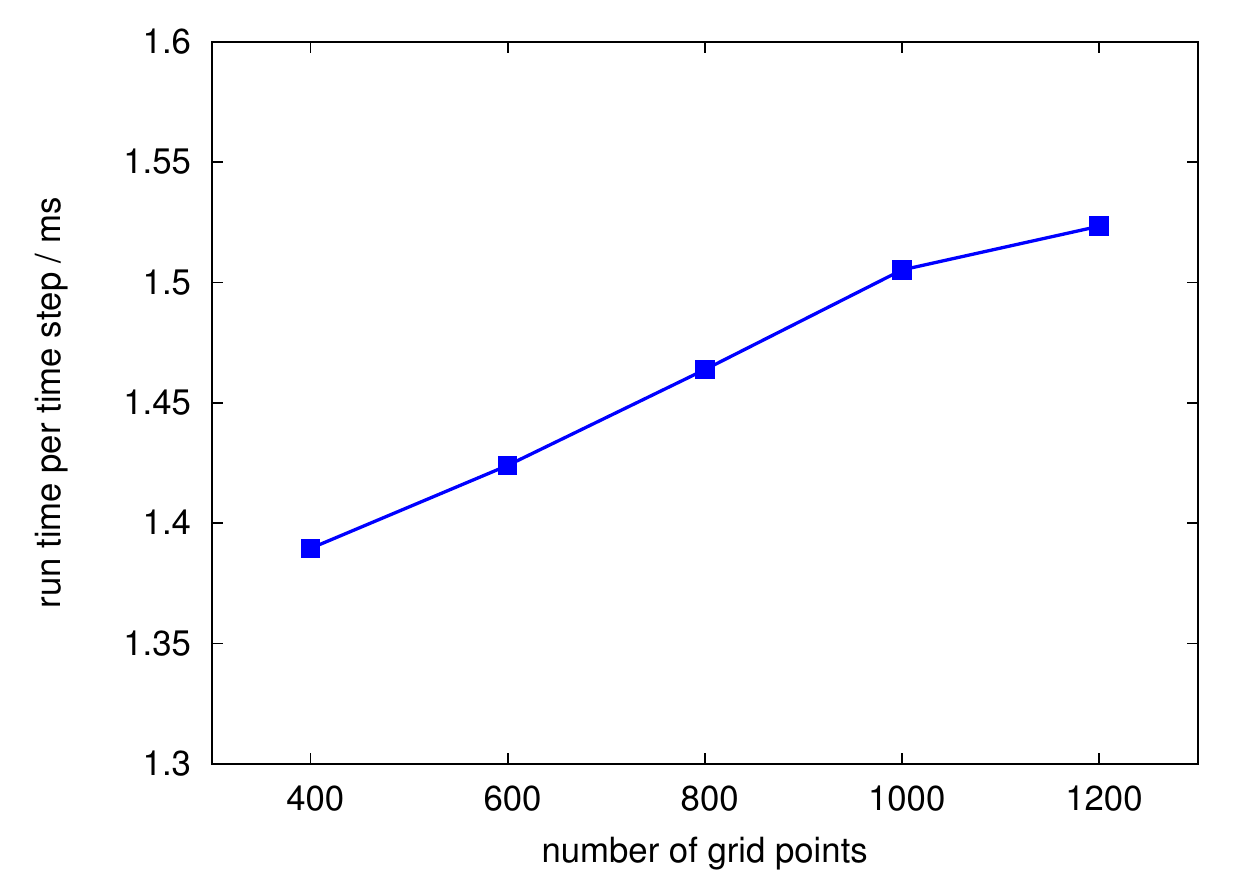}
  \end{center}
  \caption{Execution time of a single cycle for different electrical grid sizes $N_{\textrm{grid}}$. Simulation runs with 1,000,000 ions and electrons, respectively at a pressure of 10~Pa.}
  \label{fig:no_gridpoints}
\end{figure}

In this section the number of grid points $N_{\textrm{grid}}$ of the electric field lattice is changed. The total number of particles and the number of grid points per sorting cell $N_{\textrm{gc}}$ remain constant. Figure \ref{fig:no_gridpoints} shows a nearly linear dependence of the execution time on $N_{\textrm{grid}}$. By increasing $N_{\textrm{grid}}$, also the CUDA kernel grid size rises and a larger number of CUDA blocks has to be called. All dependencies are linear, small variances in figure \ref{fig:no_gridpoints} can be explained by changing number of particles per sorting cell. This behavior is different from CPU implementations, where run-time is nearly independent of the electric field grid size.

\subsection{Speed-up to CPU}
For comparison to classical CPU approaches, all diagnostics in CPU code xpdp1 \cite{xoopic_code} were removed, so only the plain PIC algorithm remains. For testing issues we use a GTX480 GPU and a single core of an Intel i7 870 CPU running at 2.93~GHz. These two devices are not only comparable with regard to up-to-dateness when establishing this study, but also in respect of their costs. For the test case of 500.000 ions and electrons (respectively) at a pressure of $20\,$ Pa, a single time step takes about $13.70\,$ms on the CPU and about $1.09\,$ms on the GPU.
\begin{figure}
  \begin{center}
    \includegraphics[width=\textwidth]{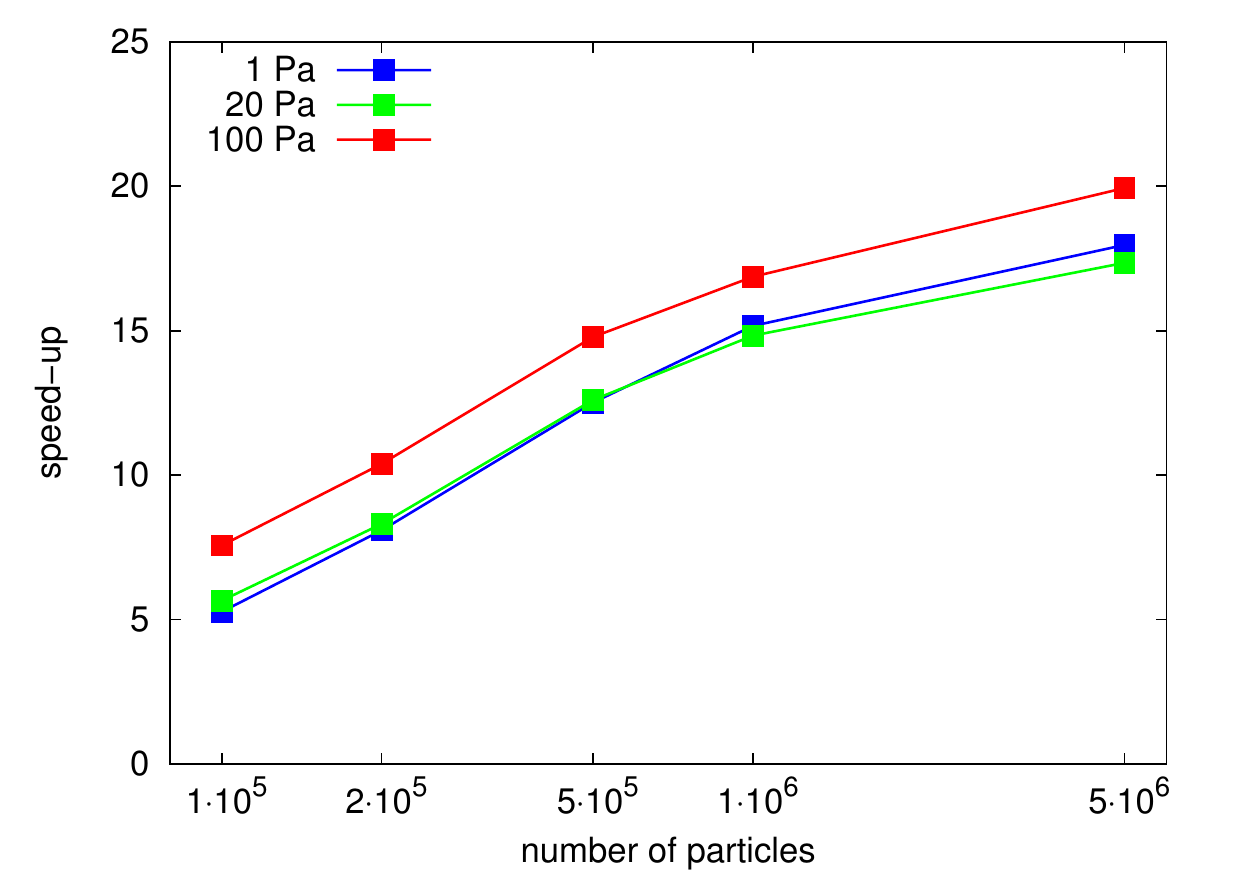}
  \end{center}
  \caption{Speed-up of the fine-sorted PIC algorithm, compared to a CPU (single core) approach as a function of the number of super-particles for different pressures. $N_{\textrm{grid}}=512$.}
  \label{fig:speedup_cpu_sorted_pressure}
\end{figure}

Figure \ref{fig:speedup_cpu_sorted_pressure} shows the speed-up of the fine-sorted GPU algorithm compared to the CPU code for three different pressures. Speed-up is calculated by dividing the CPU run-time by the GPU run-time. For increasing number of particles $N_\textrm{total}$ the system behaves as expected: low number of particles can not utilize the GPU fully, so increasing $N_\textrm{total}$ also enhances the speed-up. For lower numbers of particles (in the range of a few thousands), overhead from calling functions on the GPU and data transfer between CPU and GPU memory becomes more and more important. Hence the CPU code can become faster than the GPU for very small systems. Since PIC statistics improve with larger systems \cite{turner:033506}, we attended to larger systems in our study. Functions in figure \ref{fig:speedup_cpu_sorted_pressure} are not saturated in respect of $N_\textrm{total}$, so for very large systems even higher speed-ups than 20 can be expected.

Increasing the pressure from 1~Pa to 20~Pa slows down the GPU algorithm for high values of $N_\textrm{total}$. This is not obvious, since higher collision rates result in less branches and thereby advantages for the GPU. At a pressure of 1~Pa, Monte-Carlo collisions only take a negligible fraction of a cycle's execution time, so do not influence the speed-up. For increasing pressure, the collision module becomes more and more important. For a pressure of 20~Pa the speed-up of just the Monte-Carlo collisions is about 11.5 for a system of 1,000,000 electrons and ions, respectively. Now the total speed-up for 1~Pa (about 15.2) is higher than this value, so if collisions become more important, the whole speed-up decreases.

An increase in pressure from 20~Pa to 100~Pa makes the simulation slightly faster. Again, the collisions become more important regarding the run-time of a cycle, but this time the collision module is more than 19 times faster than the 1~Pa case. This results in an overall increase in speed.

As on the CPU, most of the time is spend for all particle related tasks, pusher, charge density assignment and for high pressure also the Monte-Carlo module. Fine-sorting of particles of course also takes a considerable amount of time. 

\section{Summary and Conclusion}
A fine-sorting particle-in-cell code, running on a single graphics processing unit was implemented, using Nvidia's CUDA environment. For testing we used newest generation GPU and CPU. All in all for a normal range of input parameters, speed-ups of about 10-20 compared to a classical CPU code could be observed. A major advantage of the fine-sorting algorithm is that it allows for a straightforward implementation of binary collisions.

The code is not bounded to any special (small) grid size. Thus, algorithms can be used for two-dimensional PIC codes, except for the field solving. Due to large numbers of particles and an efficient two-dimensional field solver, much higher speed-ups than in the one-dimensional case can be expected.

\section{Acknowledgment}
Financial support for P M from the Research School of the Ruhr-University Bochum and for D E from Deutsche Forschungsgemeinschaft within the SFB-TR~87 and FOR~1123 is gratefully acknowledged. We thank the Plasma Theory and Simulation Group in Berkeley for providing access to their code. We particularly thank Prof. A Z Panagiotopoulos as well as Dr. S Barr from Princeton university chemical engineering for sharing their GPU system.

\bibliographystyle{unsrt}
\bibliography{references}

\begin{thebibliography}{10}

\bibitem{hoomd}
Joshua~A. Anderson, Chris~D. Lorenz, and A.~Travesset.
\newblock General purpose molecular dynamics simulations fully implemented on
  graphics processing units.
\newblock {\em Journal of Computational Physics}, 227(10):5342 -- 5359, 2008.

\bibitem{badal:4878}
Andreu Badal and Aldo Badano.
\newblock Accelerating monte carlo simulations of photon transport in a
  voxelized geometry using a massively parallel graphics processing unit.
\newblock {\em Medical Physics}, 36(11):4878--4880, 2009.

\bibitem{kersevan:1017}
R.~Kersevan and J.-L. Pons.
\newblock Introduction to molflow+: New graphical processing unit-based monte
  carlo code for simulating molecular flows and for calculating angular
  coefficients in the compute unified device architecture environment.
\newblock {\em J. Vac. Sci. Technol.}, 27(4):1017--1023, 2009.

\bibitem{Anderson2007298}
Amos~G. Anderson, William A.~Goddard III, and Peter Schröder.
\newblock Quantum monte carlo on graphical processing units.
\newblock {\em Computer Physics Communications}, 177(3):298 -- 306, 2007.

\bibitem{DBD_paper_1}
Priyadarshini Rajasekaran, Philipp Mertmann, Nikita Bibinov, Dirk Wandke,
  Wolfgang Viöl, and Peter Awakowicz.
\newblock Dbd plasma source operated in single-filamentary mode for therapeutic
  use in dermatology.
\newblock {\em Journal of Physics D: Applied Physics}, 42(22):225201 (8pp),
  2009.

\bibitem{doi:10.1021/ct900275y}
M.~J. Harvey and G.~De~Fabritiis.
\newblock An implementation of the smooth particle mesh ewald method on gpu
  hardware.
\newblock {\em Journal of Chemical Theory and Computation}, 5(9):2371--2377,
  2009.

\bibitem{doi:10.1021/ct9005079}
Andrey Asadchev, Veerendra Allada, Jacob Felder, Brett~M. Bode, Mark~S. Gordon,
  and Theresa~L. Windus.
\newblock Uncontracted rys quadrature implementation of up to g functions on
  graphical processing units.
\newblock {\em Journal of Chemical Theory and Computation}, 6(3):696--704,
  2010.

\bibitem{birdsall_book}
C~K Birdsall and A~B Langdon.
\newblock {\em Plasma Physics via Computer Simulation}.
\newblock Taylor and Francis Group, LLC, 2005.

\bibitem{hockney_book}
R~W Hockney and J~W Eastwood.
\newblock {\em Computer Simulation using Particles}.
\newblock IOP Publishing Ltd, 1988.

\bibitem{tajima_book}
T~Tajima.
\newblock {\em Computational Plasma Physics}.
\newblock Westview Press, 2004.

\bibitem{turner:033506}
M~M Turner.
\newblock Kinetic properties of particle-in-cell simulations compromised by
  monte carlo collisions.
\newblock {\em Physics of Plasmas}, 13(3):033506, 2006.

\bibitem{cuda_programming_guide}
NVidia.
\newblock {\em CUDA Programming Guide}, 3.1 edition, 7 2010.

\bibitem{Stantchev20081339}
George Stantchev, William Dorland, and Nail Gumerov.
\newblock Fast parallel particle-to-grid interpolation for plasma pic
  simulations on the gpu.
\newblock {\em Journal of Parallel and Distributed Computing}, 68(10):1339 --
  1349, 2008.
\newblock General-Purpose Processing using Graphics Processing Units.

\bibitem{Tskhakaya2007829}
D.~Tskhakaya and R.~Schneider.
\newblock Optimization of pic codes by improved memory management.
\newblock {\em Journal of Computational Physics}, 225(1):829 -- 839, 2007.

\bibitem{160337}
J~P Verboncoeur, M~V Alves, V~Vahedi, and C~K Birdsall.
\newblock Simultaneous potential and circuit solution for 1d bounded plasma
  particle simulation codes.
\newblock {\em J. Comput. Phys.}, 104(2):321--328, 1993.

\bibitem{thomas_algorithm}
S~D Conte and C~De Boor.
\newblock {\em Elementary Numerical Analysis; an Algorithmic Approach}.
\newblock McGraw-Hill, New York, 1972.

\bibitem{verboncoeur_review}
J~P Verboncoeur.
\newblock Particle simulation of plasmas: review and advances.
\newblock {\em Plasma Physics and Controlled Fusion}, 47(5A):A231--A260, 2005.

\bibitem{Birdsall_paper}
C~K Birdsall.
\newblock Particle-in-cell charged-particle simulations, plus monte carlo
  collisions with neutral atoms, pic-mcc.
\newblock {\em Plasma Science, IEEE Transactions on}, 19(2):65 --85, apr 1991.

\bibitem{Marsaglia:2003:JSSOBK:v08i14}
G~Marsaglia.
\newblock Xorshift rngs.
\newblock {\em Journal of Statistical Software}, 8(14):1--6, 7 2003.

\bibitem{Phelps_ionen_cross_sections}
A~V Phelps.
\newblock Cross sections and swarm coefficients for nitrogen ions and neutrals
  in n2 and argon ions and neutrals in ar for energies from 0.1 ev to 10 kev.
\newblock {\em Journal of Physical and Chemical Reference Data}, 20:557--573,
  May 1991.

\bibitem{phelps:747}
A.~V. Phelps.
\newblock The application of scattering cross sections to ion flux models in
  discharge sheaths.
\newblock {\em Journal of Applied Physics}, 76(2):747--753, 1994.

\bibitem{argon_electronen_cross_sections}
A~Yanguas-Gil, J~C, and L~L Alves.
\newblock An update of argon inelastic cross sections for plasma discharges.
\newblock {\em Journal of Physics D: Applied Physics}, 38(10):1588, 2005.

\bibitem{xoopic_code}
The plasma theory and simulation group - ptsg software, xpdp1, xpdp2, xpds1,
  xoopic, http://ptsg.eecs.berkeley.edu/.

\end{thebibliography}

\clearpage

\end{document}